\documentclass[draftclsnofoot, onecolumn, comsoc, 12pt]{IEEEtran} 

\ifCLASSINFOpdf
\else
   \usepackage[dvips]{graphicx}
\fi
\usepackage{url}
\usepackage{hhline}
\usepackage{graphicx}
\usepackage{multicol}
\usepackage{graphicx}
\usepackage{amssymb}
\usepackage{amsmath}
\DeclareMathOperator*{\argmax}{arg\,max}

\usepackage{calc}
\usepackage{array}
\usepackage{epstopdf}
\usepackage{mathtools}
\usepackage{mathrsfs}
\usepackage{bm}
\usepackage{cite}
\usepackage{boldline}
\usepackage{amsthm}
\usepackage{algorithm}
\usepackage{algorithmicx, algpseudocode}
\usepackage{amsthm}

\newtheorem{remark}{Remark}

\usepackage{amsmath}
\usepackage{mathrsfs}
\usepackage{amsmath}
\usepackage[short,c1,nocomma]{optidef}

\usepackage{booktabs,makecell,tabularx}

\newcolumntype{C}[1]{>{\centering\arraybackslash}p{#1}}
\newcolumntype{L}{>{\raggedright\arraybackslash}X}

\usepackage{siunitx}
\usepackage{etoolbox}
\newrobustcmd{\B}{\bfseries}

\usepackage{comment}
\usepackage{algorithm}
\usepackage{algpseudocode}
\usepackage{enumitem} 

\usepackage{colortbl}     
\usepackage{subfigure}
\usepackage{color, soul}
\usepackage{stfloats}
\usepackage{float}
\usepackage{multirow}
\usepackage{wrapfig}
\usepackage{booktabs}
\usepackage{slashbox}
\definecolor{LightBlue}{rgb}{0.75,0.936,1.00}
\sethlcolor{LightBlue}
\definecolor{LightCyan}{rgb}{0.88,1,1}

\usepackage{xcolor}

\begin{document}
\bstctlcite{IEEEexample:BSTcontrol}
\title{Space-Time Rate-Splitting Multiple Access \\for Multibeam LEO Satellite Networks
}
\author{Jaehyup Seong, 
Byungju Lee, 
Aryan Kaushik
, and Wonjae Shin
    \thanks{ Jaehyup Seong and Wonjae Shin are with the School of Electrical Engineering, 
    Korea University, Seoul 02841, South Korea 
    (email: jaehyup@korea.ac.kr; wjshin@korea.ac.kr);
    Byungju Lee is with the Department of Information and Telecommunication Engineering, 
    Incheon National University, Incheon 22012, South Korea (e-mail: {bjlee@inu.ac.kr});
    Aryan Kaushik  is with the Department of Computing and Mathematics, 
    Manchester Metropolitan University, Manchester M1 5GD, United Kingdom (e-mail: {a.kaushik@mmu.ac.uk}).
    }}
\maketitle

\begin{abstract} 
This paper proposes a novel space-time rate-splitting multiple access (ST-RSMA) framework for multibeam low Earth orbit (LEO) satellite communications (SATCOM) systems, {where space-time coding is integrated into the common stream transmission. This design enables \textit{full diversity gain} in the common stream transmission for all users, regardless of the uncertainty of the channel state information (CSI) and network load conditions, thereby overcoming the performance limitations of conventional RSMA that employs a single beamforming vector for all users.} 
{To further enhance performance, we develop a weighted minimum mean square error (WMMSE)-based algorithm tailored to ST-RSMA that jointly optimizes the power allocation for the common stream and the power/beamforming vectors for private streams, aiming to maximize the minimum user rate.}
Numerical results show that ST-RSMA significantly outperforms conventional RSMA and other multiple access techniques, offering a robust and scalable solution for LEO SATCOM.
\end{abstract}
\section{Introduction}
Multibeam low Earth orbit (LEO) satellite communications (SATCOM) is gaining prominence for providing high-throughput and massive connectivity across vast regions with low latency, making it a cornerstone of fifth-generation (5G) and beyond wireless networks. Expanding coverage, where terrestrial networks fall short, multibeam LEO SATCOM enables a broadband service with truly ubiquitous connectivity \cite{perez2019signal, jamshed2025tutorial}.


Despite its promise, providing reliable high-speed data services to a large number of users across a wide coverage area presents significant technical challenges.
Specifically, the scarcity of spectral resources necessitates aggressive resource reuse to achieve high throughput. However, this approach inevitably leads to severe intra-beam and inter-beam interference. {Managing such interference requires accurate channel state information (CSI); however, the high mobility of LEO satellites (traveling at approximately \num{7.56} km/s at an altitude of \num{600} km) and the significant round-trip delay (RTD) (approximately \num{25.77} ms) pose substantial challenges in acquiring precise CSI \cite{3gpp_ntn, you2020massive, 10844052}.}
Further, the high user density within extensive service zones ($\num{100}-\num{1000}$ km) exacerbates network congestion, where available spatial degrees of freedom are insufficient to effectively mitigate interference \cite{10559954}. 

Rate-splitting multiple access (RSMA) has recently stood out as a promising multiple access (MA) technique that enhances robustness to imperfect CSI at the transmitter (CSIT) while effectively accommodating a large number of users.  
The core of RSMA relies on adaptive message splitting, where each user's message is divided into a common part, shared across users, and a private part, intended for individual decoding. 
Through dynamic power allocation between common and private parts according to channel conditions and network load, RSMA achieves robustness against imperfect CSIT while maintaining scalability and high spectral efficiency \cite{park2023rate}.


Building on the strengths of RSMA, extensive research in multibeam SATCOM has demonstrated its superiority over conventional MA, including spatial division MA (SDMA) and non-orthogonal MA (NOMA) \cite{yin2020rate_J, cui2023energy}. By jointly optimizing power allocation and the beamforming vector for the common/private streams, RSMA has unlocked higher spectral and energy efficiencies in user-overloaded multibeam SATCOM with imperfect CSI. Beyond mitigating intra-network interference, RSMA's flexible use of the common stream has also been explored to manage inter-network interference in geostationary orbit (GEO)-LEO satellite coexistence networks and integrated satellite-terrestrial networks (ISTN) \cite{10266774, 10636955}.


{Nonetheless, the achievable rate of the common stream is fundamentally constrained by the requirement that all users must successfully decode it. 
As user density increases in LEO satellite networks, enhancing the diversity gain of the common stream becomes essential for RSMA to scale effectively. However, relying on a single common beamforming vector restricts the system’s ability to provide sufficient diversity gain to all users, particularly in densely populated networks. This limitation becomes more critical under severe CSIT inaccuracies, where the beamformer is optimized based on imprecise CSI at the satellite. Moreover, in practical systems with finite block lengths, insufficient diversity gain can result in decoding failures of the common stream at some users, which leads to imperfect successive interference cancellation (SIC) and severely degrades the overall RSMA performance.}


{To overcome these limitations,} in this paper, we introduce a space-time RSMA (ST-RSMA) framework for LEO SATCOM that integrates space-time coding \cite{alamouti1998simple} into the transmission of the common stream. {This approach \textit{eliminates the need for a dedicated beamforming vector} and \textit{ensures full diversity gain} in the common stream transmission to all users with low complexity, regardless of the network load and CSI uncertainty.}
{The main contributions are summarized as follows:}
\begin{itemize} 
\item We put forth a new ST-RSMA framework for LEO SATCOM with high user density, where accurate CSI acquisition is particularly challenging due to the large RTD and rapid satellite motion.
{By incorporating space-time coding into the common stream transmission, our approach achieves full diversity gain for all users without relying on a dedicated beamforming vector for the common stream, regardless of CSI accuracy and user density.}
\item {To guarantee fairness across the coverage area under imperfect CSIT}, we formulate a max-min fairness problem based on the proposed ST-RSMA framework. {To address its non-convexity, we develop a computationally efficient weighted minimum mean square error (WMMSE)-based algorithm tailored to ST-RSMA, jointly optimizing the power allocation for common and private streams, as well as the private beamforming vectors.}
\item {Extensive numerical evaluations based on key parameters from the 3GPP non-terrestrial networks (NTN) standards \cite{3gpp_ntn} show that the proposed ST-RSMA significantly improves the minimum spectral efficiency in LEO SATCOM with imperfect CSIT, consistently outperforming conventional RSMA and other state-of-the-art MA techniques.}
\end{itemize}

{ 
\textit{Notations:} Throughout the paper, standard letters, lowercase boldface letters, and uppercase boldface letters denote scalars, vectors, and matrices, respectively.
The notations $t$, $(\cdot)^{\sf{T}}$, $(\cdot)^{\sf{H}}$, $(\cdot)^{\sf{*}}$, and $\mathbb{E}[\cdot]$ represent the time index, transpose, hermitian, complex conjugate, and statistical expectation, respectively. }

\section{System Model and Problem Formulation} 
We consider a multibeam LEO SATCOM system where an $N_{\sf{t}}$-antenna LEO satellite serves $K$ single-antenna users using a single-feed-per-beam  architecture \cite{perez2019signal}.\footnote{{While this work focuses on a single-feed-per-beam architecture, commonly used to maximize coverage with a limited number of satellite transmit feeds, the proposed ST-RSMA scheme can be readily adopted to multi-feed-per-beam configurations.}} The satellite-to-$k$-th user channel vector $\mathbf{h}_{k} \in \mathbb{C}^{N_{\sf{t}} \times 1}$ consists of the satellite beam radiation pattern, free-space path loss, user antenna gain, and signal phase components \cite{yin2020rate_J}.
Thus, the $n_{\sf{t}}$-th element of $\mathbf{h}_k$, representing the $n_{\sf{t}}$-th feed-to-$k$-th user channel, is given by { $h_{k,n_{\sf{t}}} = \frac{\sqrt{G_{k, n_{\sf{t}}} G_{\sf{Rx}}}}{4\pi {d_k}/\lambda_{\sf{c}} \sqrt{\kappa {T_{\sf sys}} {B}}} e^{-j\phi_{k,n_{\sf{t}}}}$},
where $G_{\sf{Rx}}$, $d_k$, $\lambda_{\sf{c}}$, $\kappa$, $T_{\sf sys}$, and $B$ denote the user terminal antenna gain, satellite-to-$k$-th user distance, carrier wavelength, Boltzmann constant, receiver noise temperature, and bandwidth, respectively.
The beam gain from the $n_{\sf{t}}$-th feed to the $k$-th user, denoted as $G_{k,n_{\sf{t}}}$, is expressed by the first- and third-order Bessel functions of the first kind, $J_1(\cdot)$ and $J_3(\cdot)$, as
{ $ G_{k,n_{\sf{t}}} = G_{\sf{Tx}}^{\sf{max}} \big[\frac{J_{1}(\mu_{k,n_{\sf{t}}})}{2\mu_{k,n_{\sf{t}}}} + 36 \frac{J_{3}(\mu_{k,n_{\sf{t}}})}{\mu_{k,n_{\sf{t}}}^{3}}\big]^{2}$}, where
$G_{\sf{Tx}}^{\sf{max}}$ is the maximum transmit antenna gain.
$\mu_{k,n_{\sf{t}}}$ is given by { $\mu_{k,n_{\sf{t}}} = 2.07123\big(\frac{\sin(\theta_{k,n_{\sf{t}}})}{\sin(\theta_{\textrm{3dB}})}\big)$}, 
where $\theta_{k,n_{\sf{t}}}$ and $\theta_{\textrm{3dB}}$ denote the angle between the $n_{\sf{t}}$-th feed's beam center and the $k$-th user and the 3 dB beamwidth, respectively.
$\phi_{k,n_{\sf{t}}}$ is the $n_{\sf{t}}$-th feed-to-$k$-th user phase component, following an independent uniform distribution over $[0, 2\pi]$.

{While the high velocity of LEO satellites induces significant Doppler shifts that result in substantially outdated CSIT, it can be partially mitigated using Doppler pre-compensation techniques at the satellite payload \cite{3gpp_ntn}. However, residual Doppler shifts, long propagation delays, and rapid satellite motion collectively result in an imperfect CSIT. Consequently, we model the estimated channel at the LEO satellite as $\hat{\mathbf{h}}_{k} = \mathbf{h}_{k} - \mathbf{e}_{k}$, where $\mathbf{e}_{k}$ denotes the CSI error that follows a complex Gaussian distribution such that $\mathbf{e}_{k}\sim \mathcal{CN}(\mathbf{0},\mathbf{\Phi}_{k})$.}

\begin{figure}[!t]
\centering
 		\includegraphics[width=0.9\linewidth]{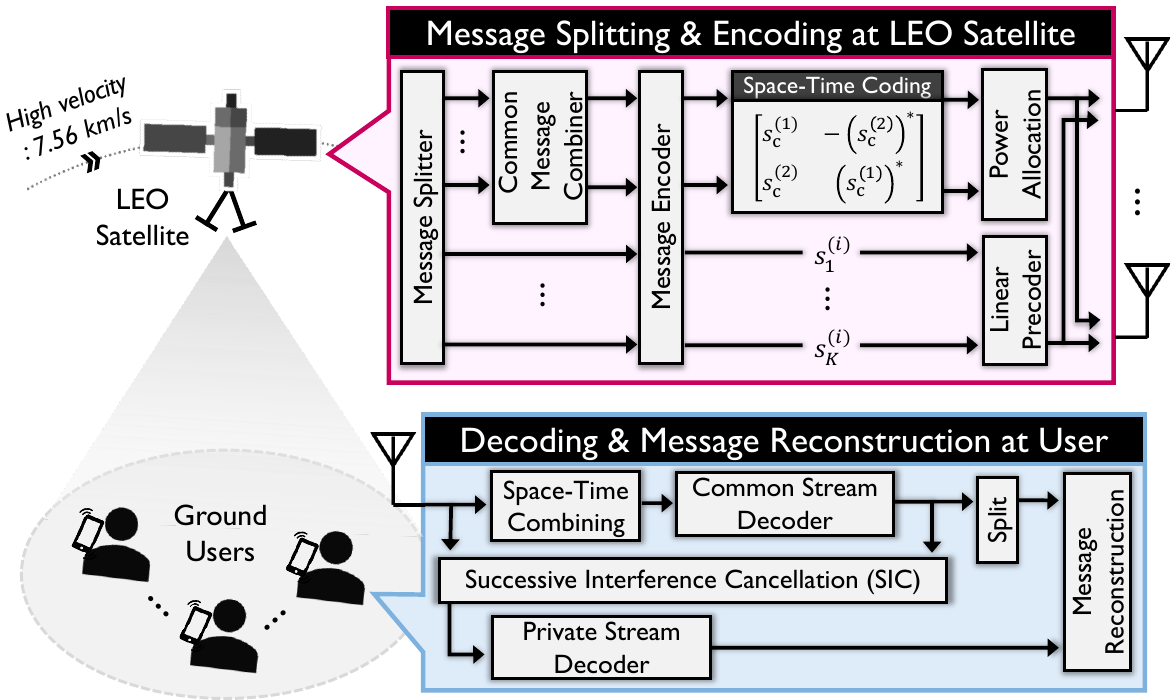}
 		\caption{System model of the proposed ST-RSMA scheme for LEO SATCOM.}
    	\label{Fig1}
\end{figure}

Employing ST-RSMA, as shown in Fig. \ref{Fig1}, the LEO satellite splits messages, denoted as $\{M_{1}^{(i)}, \cdots, M_{K}^{(i)}\}$, $\forall i \in \{1, 2\}$, into common and private parts as
$M_{k}^{(i)} \rightarrow \{M_{{\sf{c}}, k}^{(i)}, M_{{\sf{p}}, k}^{(i)} \}, \forall k \in \mathcal{K} \triangleq \{ 1,\cdots, K \}$.
For each $i \in \{1, 2\}$, the common messages $\{M_{{\sf{c}},1}^{(i)}, \cdots, M_{{\sf{c}},K}^{(i)}\}$ are aggregated into a single common message $M_{\sf{c}}^{(i)}$. Then, $M_{{\sf{c}}}^{(1)}$ and $M_{{\sf{c}}}^{(2)}$ are respectively encoded into common streams $s_{{\sf{c}}}^{(1)}$ and $s_{{\sf{c}}}^{(2)}$ using a shared codebook among all users, and then mapped via space-time coding as
{
\begin{align}
\mathbf{s}_{{\sf{c}}}^{(1)} = \big[s_{{\sf{c}}}^{(1)}, s_{ {\sf{c}}}^{(2)} \big]^{\sf{T}}, \,\, \mathbf{s}_{{\sf{c}}}^{(2)} = \big[ -(s_{{\sf{c}}}^{(2)})^{*}, (s_{{\sf{c}}}^{(1)})^{*} \big]^{\sf{T}}.
\end{align}  
}Each user's private message is encoded into a private stream using a codebook, known only to the corresponding user, forming the private streams $\{s_{1}^{(i)}, \cdots, s_{K}^{(i)}\}$, and then the private streams are linearly combined using private precoding vectors $\mathbf{p}_k \in \mathbb{C}^{N_{\sf{t}} \times 1}$, $\forall k \in \mathcal{K}$, such as
$ \sum_{j=1}^{K} \mathbf{p}_{j} s_{j}^{(i)}$, $\forall i \in \{1, 2\}$. The common and private streams are drawn from an independent and identical distribution (i.i.d.) such that $s_{\sf{c}}^{(i)}, s_{k}^{(i)} \sim \mathcal{CN}{(0,1)}$.

Then, the LEO satellite sequentially transmits a superimposed signal over two symbol periods. To elaborate on this process, we first consider the case where the LEO satellite is equipped with two transmit antenna feeds. The transmitted signals at the first and second symbol periods are given by 
{
\begin{align}
\mathbf{x}(t) \!=\! \sqrt{\!\frac{P_{\sf{c}}}{2}} \mathbf{s}_{ {\sf{c}}}^{(1)} \!\!+\! \sum_{j=1}^{K} \mathbf{p}_{j} s_{j}^{(1)}\!\!, \,\, \mathbf{x}(t+T) \!=\! \sqrt{\!\frac{P_{\sf{c}}}{2}} \mathbf{s}_{{\sf{c}}}^{(2)} \!\!+\! \sum_{j=1}^{K} \mathbf{p}_{j} s_{j}^{(2)}\!\!, 
\end{align} 
}where $T$ and $P_{\sf{c}}$ are the symbol duration and allocated power for common streams, respectively. 
Therefore, each user receives two
signals over consecutive periods with the $k$-th user’s received signals for the first and second periods, given by
{
\begin{align}
y_{k}(t) \!=\! \mathbf{h}_{k}^{\mathsf{H}} \mathbf{x}(t) 
    \!+\!  n_{k}(t), \, y_{k}(t+T) \!=\! \mathbf{h}_{k}^{\mathsf{H}} \mathbf{x}(t+T) 
    \!+\!  n_{k}(t+T), 
\end{align}
}where $n_{k}$ is the additive white Gaussian noise (AWGN) that follows i.i.d. such that $n_k\sim\mathcal{CN}{(0,\sigma_{\sf{n}}^2)}$, $\forall k \in \mathcal{K}$. Thereafter, the $k$-th user constructs a received signal vector as
{
\begin{align}
\label{stacked_received_signal}
    \begin{bmatrix} 
        y_{k}(t) \\
        y_{k}^{*}(t+T) 
    \end{bmatrix}  & =   
    \begin{bmatrix} 
        h^{*}_{k,1} & h^{*}_{k,2} \\
        h_{k,2} & -h_{k,1} 
    \end{bmatrix}
    \begin{bmatrix} 
        \sqrt{P_{\sf{c}}/2} \cdot s_{{\sf{c}}}^{(1)}\\
        \sqrt{P_{\sf{c}}/2}  \cdot s_{{\sf{c}}}^{(2)} 
    \end{bmatrix} \nonumber \\
    & + \begin{bmatrix} 
        \mathbf{h}_{k}^{\mathsf{H}}\sum_{j=1}^{K} \mathbf{p}_{j} s_{j}^{(1)} + n_{k}(t) \\
        \big(\mathbf{h}_{k}^{\mathsf{H}}\sum_{j=1}^{K} \mathbf{p}_{j} s_{j}^{(2)} + n_{k}(t+T)\big)^{*}
    \end{bmatrix}.
\end{align}
}Notably, since the channel matrix in the equation (\ref{stacked_received_signal}) exhibits an orthogonal structure, the common streams $s_{\sf{c}}^{(1)}$ and $s_{\sf{c}}^{(2)}$ can be efficiently separated using combining technique as follows:
{
\begin{align}
\label{combining}
        & \!\!\! \begin{bmatrix} 
        h_{k,1} & h^{*}_{k,2} \\
        h_{k,2} & -h^{*}_{k,1} 
    \end{bmatrix} \begin{bmatrix} 
        y_{k}(t) \\
        y_{k}^{*}(t+T) 
    \end{bmatrix}  =   
    \begin{bmatrix} 
        \Vert \mathbf{h}_{k} \Vert^{2} & \!\!\!\!\!\!\!\!\! 0 \\
        0 & \!\!\!\!\!\!\!\!\! \Vert \mathbf{h}_{k} \Vert^{2} 
    \end{bmatrix}
    \begin{bmatrix} 
        \sqrt{P_{\sf{c}}/2}  \cdot s_{{\sf{c}}}^{(1)}\\
        \sqrt{P_{\sf{c}}/2}  \cdot s_{{\sf{c}}}^{(2)} 
    \end{bmatrix} \nonumber \\
    & \!\!\! + \begin{bmatrix} 
        h_{k,1} & h^{*}_{k,2} \\
        h_{k,2} & -h^{*}_{k,1} 
    \end{bmatrix} 
    \begin{bmatrix} 
        \mathbf{h}_{k}^{\mathsf{H}}\sum_{j=1}^{K} \mathbf{p}_{j} s_{j}^{(1)} + n_{k}(t) \\
        \big(\mathbf{h}_{k}^{\mathsf{H}}\sum_{j=1}^{K} \mathbf{p}_{j} s_{j}^{(2)} + n_{k}(t+T)\big)^{*}
    \end{bmatrix}.
\end{align}
}{Following this, the common streams $s_{\sf{c}}^{(1)}$ and $s_{\sf{c}}^{(2)}$
in each vector element of the equation (\ref{combining}) are first decoded while treating the private streams present in the corresponding vector element as noise. Therefore, $h_{k,1}(\mathbf{h}_{k}^{\mathsf{H}}\sum_{j=1}^{K} \mathbf{p}_{j} s_{j}^{(1)}) + h_{k,2}^{*}(\mathbf{h}_{k}^{\mathsf{H}}\sum_{j=1}^{K} \mathbf{p}_{j} s_{j}^{(2)})+ h_{k,1}n_k(t) + h_{k,2}^{*}n_k(t+T)^{*}$ is treated as noise when decoding $s_{\sf{c}}^{(1)}$, while $h_{k,2}(\mathbf{h}_{k}^{\mathsf{H}}\sum_{j=1}^{K} \mathbf{p}_{j} s_{j}^{(1)}) - h_{k,1}^{*}(\mathbf{h}_{k}^{\mathsf{H}}\sum_{j=1}^{K} \mathbf{p}_{j} s_{j}^{(2)}) + h_{k,2}n_k(t) - h_{k,1}^{*}n_k(t+T)^{*}$ is treated as noise when decoding $s_{\sf{c}}^{(2)}$.}
{After decoding the common streams, $s_{\sf{c}}^{(1)}$ and $s_{\sf{c}}^{(2)}$ are re-encoded and removed from $y_{k}(t)$ and $y_{k}(t+T)$ through the SIC technique as $y_{k, {\sf{SIC}}}(t) = \sum_{j=1}^{K} \mathbf{h}_{k}^{\mathsf{H}} \mathbf{p}_{j} s_{j}^{(1)} 
    +  n_{k}(t)$ and $y_{k, {\sf{SIC}}}(t+T) = \sum_{j=1}^{K} \mathbf{h}_{k}^{\mathsf{H}} \mathbf{p}_{j} s_{j}^{(2)}  
    +  n_{k}(t+T)$, respectively.}
{Subsequently, $s_{k}^{(1)}$ and $s_{k}^{(2)}$ are decoded separately, each treating the other users' private streams as noise. Thus, $\sum_{j=1, j \neq k}^{K} \mathbf{h}_{k}^{\mathsf{H}} \mathbf{p}_{j} s_{j}^{(1)} + n_k(t)$ is treated as noise when decoding $s_{k}^{(1)}$, while $\sum_{j=1, j \neq k}^{K} \mathbf{h}_{k}^{\mathsf{H}} \mathbf{p}_{j} s_{j}^{(2)} + n_k(t+T)$ is treated as noise when decoding $s_{k}^{(2)}$.}
Therefore, common and private spectral efficiencies for the $k$-th user are derived as 
{
\begin{align}
    \label{common rate_tot}
     & {R}_{{\sf{c}},k} \!=\! \sum_{i=1}^{2}{R}_{{\sf{c}},k}^{(i)} = \log_2\left(1+\frac{\Vert\mathbf{h}_{k}\Vert^2 \frac{P_{\sf{c}}}{2}}{\sum_{j=1}^{K}\vert\mathbf{h}_{k}^{\sf{H}}\mathbf{p}_{j}\vert^{2} + \sigma_{{\sf{n}}}^{2}}\right),  \\
    \label{private rate_tot}
     & {R}_{{\sf{p}},k} \!=\! \sum_{i=1}^{2}{R}_{{\sf{p}},k}^{(i)} = \log_2\left(\!1\!+\!\frac{\vert\mathbf{h}_{k}^{\sf{H}}\mathbf{p}_{k}\vert^{2}}{\sum_{j=1, j \neq k }^{K}\vert\mathbf{h}_{k}^{\sf{H}}\mathbf{p}_{j}\vert^{2} \!+\! \sigma_{{\sf{n}}}^{2}}\!\right)\!.
\end{align}}

To design a robust space-time RSMA-based precoder under imperfect CSIT, we characterize common and private ergodic spectral efficiencies as
$\mathbb{E}_{\{\mathbf{e}_{k}\}}\left[{R}_{{\sf{c}},k} \vert\hat{\mathbf{h}}_{k} \right]$ and $\mathbb{E}_{\{\mathbf{e}_{k}\}}\left[{R}_{{\sf{p}},k} \vert\hat{\mathbf{h}}_{k} \right]$, $\forall k \in \mathcal{K}$.
However, since obtaining a closed-form expression for the ergodic equation is generally intractable, we use the sample average approximation (SAA) technique to approximate the common and private ergodic spectral efficiencies using an empirical average.
To this end, we generate $S$ samples of $\mathbf{h}_{k}^{\left(s\right)}$ as $\mathbf{h}_{k}^{\left(s\right)} = \hat{\mathbf{h}}_{k} + \mathbf{e}_{k}^{\left(s\right)}$, where $\hat{\mathbf{h}}_{k}$ is given at the LEO satellite and $\mathbf{e}_{k}^{\left(s\right)}$ is randomly generated, i.e., $\mathbf{e}_{k}^{(s)}\sim \mathcal{CN}(\mathbf{0},\mathbf{\Phi}_{k})$, $\forall s \in \mathcal{S} \triangleq \{1,\cdots,S\}$. 
Then, the empirical averages of the common and private spectral efficiencies are computed as 
{
\begin{align}
    \label{empirical average common rate}
    &\mathbb{E}_{\{\mathbf{e}_{k}\}}\left[{R}_{{\sf{c}},k} \vert\hat{\mathbf{h}}_{k} \right] \overset{(a)}{\approx} \bar{R}_{{\sf{c}},k} \triangleq \frac{1}{S}\sum\limits_{s=1}^{S} R_{{\sf{c}},k}(\mathbf{h}_{k}^{\left(s\right)}), \\
    \label{empirical average private rate}
    &\mathbb{E}_{\{\mathbf{e}_{k}\}}\left[{R}_{{\sf{p}},k} \vert\hat{\mathbf{h}}_{k} \right] \overset{(b)}{\approx} \bar{R}_{{\sf{p}},k} \triangleq \frac{1}{S}\sum\limits_{s=1}^{S} R_{{\sf{p}},k}(\mathbf{h}_{k}^{\left(s\right)}).
\end{align}
}Herein, $R_{{\sf{c}},k}(\mathbf{h}_{k}^{\left(s\right)})$ and $R_{{\sf{p}},k}(\mathbf{h}_{k}^{\left(s\right)})$ are the common and private spectral efficiencies for the realized channel sample $\mathbf{h}_{k}^{\left(s\right)}$, and the approximations $(a)$ and $(b)$ become tight as $S \rightarrow \infty$.

Our goal is to maximize the minimum user spectral efficiency within the satellite coverage area under imperfect CSIT. This can be expressed as the following optimization problem.
\begingroup
\let\oldbodyobjLong\bodyobjLong
\renewcommand{\bodyobjLong}[4]{%
  \ifthenelse{\isempty{#4}}{%
    &\text{\normalsize$\mathscr{P}_1:$}~\operatorname*{maximize}_{\displaystyle #1}\ #2\nonumber\span\span\span\span%
  }{%
    #4~&\text{\normalsize$\mathscr{P}_1:$}~\operatorname*{maximize}_{\displaystyle #1}\ #2\nonumber\span\span\span\span%
  }%
}
\begin{maxi!}|l|[2]
  {_{P_{\sf c},\, \mathbf{P},\, \mathbf{c}}}
  {\min\limits_{k \in \mathcal{K}}\!\big(\bar{R}_{{\sf p},k}+C_k\big)}
  {}{} 
  \addConstraint{\bar{R}_{{\sf c},k} \ge \sum_{j=1}^{K}C_j, ~ \forall k\in\mathcal{K},}\label{PF1CST1}
  \addConstraint{C_k \ge 0, ~ \forall k\in\mathcal{K},}\label{PF1CST2}
  \addConstraint{P_{\sf c}+\sum_{j=1}^{K}\lVert \mathbf{p}_j\rVert^2 \le P_{\sf t}, ~ P_{\sf c}\ge 0,}\label{PF1CST3}
\end{maxi!}
\let\bodyobjLong\oldbodyobjLong
\endgroup
where $P_{\sf{c}}$ denotes the transmit power allocated to the common streams, and $C_{k}$ represents the portion of $k$-th user from the common spectral efficiency.  
$\mathbf{P}=[ \mathbf{p}_1,\cdots,\mathbf{p}_K]\in \mathbb{C}^{{N_{\sf{t}}}\times{K}}$ denotes private precoding matrix while $\mathbf{c}=[C_1,\cdots, C_K]^{\sf{T}}$ represents the vector composed of common portions assigned to each user. The constraint (\ref{PF1CST1}) guarantees that the common stream remains decodable for all users, and constraint (\ref{PF1CST2}) ensures the non-negativity of the common portions. The constraint (\ref{PF1CST3}) imposes a total transmit power constraint, where $P_{\sf{t}}$ is the total transmit power budget for the LEO satellite.

{Note that space-time coding for the common stream transmission in ST-RSMA requires only lightweight onboard operations, including complex conjugation and sign inversion across two symbol periods. These linear operations can be efficiently implemented using standard DSPs or FPGAs, deployed in satellite payloads. Importantly, ST-RSMA alleviates the computational burden on resource-constrained satellite onboard by eliminating the need for common beamforming vector design.}


{\begin{remark} 
{\rm \textbf{(Feasibility analysis of ST-RSMA under 3GPP NTN standards)}: 
Due to the dominant line-of-sight characteristic of satellite channels, satellite networks employ orthogonal frequency division multiplexing (OFDM) subcarrier spacings of \num{60} and \num{120} kHz for Ka-band operations \cite{3gpp_ntn}, corresponding to symbol durations of \num{16.67} and \num{8.33} $\mu\text{s}$, respectively.
Given that the length of cyclic prefix (CP) overhead in OFDM systems is typically set to around 7 $\%$ of the symbol durations \cite{dahlman20205g}, the total OFDM symbol durations including CP become \num{17.84} and \num{8.91} $\mu\text{s}$, respectively. Furthermore, with Doppler pre-compensation technique applied at the LEO satellite payload, as specified in \cite{3gpp_ntn}, the maximum residual Doppler shifts for beam radius of \num{10} and \num{25} km are approximately \num{8.4} and \num{21} kHz at \num{20} GHz carrier frequency, yielding coherence times of about \num{50.37} and \num{20.15} $\mu\text{s}$, respectively, under Clarke’s model. This implies that, for a \num{10} km beam radius, the coherence time exceeds the duration of 
two consecutive OFDM symbols for both  \num{60} and \num{120} kHz subcarrier spacings. Even for a \num{25} km beam radius, the \num{120} kHz configuration still satisfies the coherence requirement.
Therefore, ST-RSMA can be reliably operated in practical multibeam LEO SATCOM scenarios.}
\end{remark}}
{\begin{remark} 
{\rm \textbf{(Extension to space-frequency RSMA for rapidly varying channels)}:
Although this work focuses on achieving full diversity gain for common stream transmission in the space-time domain, the same principle can be extended to the space-frequency domain, as in \cite{alamouti1998simple}, depending on the constraints of the system. Specifically, space-frequency RSMA, which exploits adjacent OFDM subcarriers instead of consecutive OFDM symbols, provides a viable alternative in scenarios where the channel coherence time is extremely short due to factors such as insufficient Doppler pre-compensation.} \end{remark}}

\section{Proposed Space-Time RSMA Precoder Design} \label{Sec3}

Since $\mathscr{P}_1$ is a non-convex problem, solving it directly poses significant challenges. To reformulate it into a tractable form, we introduce auxiliary variables $\alpha_{{\sf{p}},k}$ and $q$ with respect to $\bar{R}_{{\sf{p}},k}$ and minimum user spectral efficiency, respectively, as 
\begingroup
\let\oldbodyobjLong\bodyobjLong
\renewcommand{\bodyobjLong}[4]{%
  \ifthenelse{\isempty{#4}}{%
    &\text{\normalsize$\mathscr{P}_2:$}~\operatorname*{maximize}_{\displaystyle #1}\ #2\nonumber\span\span\span\span%
  }{%
    #4~&\text{\normalsize$\mathscr{P}_2:$}~\operatorname*{maximize}_{\displaystyle #1}\ #2\nonumber\span\span\span\span%
  }%
}
\begin{maxi!}|l|[2]
  {_{P_{\sf c},\, \mathbf{P},\, \mathbf{c},\, \boldsymbol{\alpha}_{\sf p},\, q}}
  {q}
  {}{} 
  \addConstraint{\alpha_{{\sf p},k} + C_k \ge q,~ \forall k\in\mathcal{K},}\label{PF2CST1}
  \addConstraint{\bar{R}_{{\sf p},k} \ge \alpha_{{\sf p},k},~ \forall k\in\mathcal{K},}\label{PF2CST2}
  \addConstraint{\alpha_{{\sf p},k} \ge 0,~ \forall k\in\mathcal{K},}\label{PF2CST3}
  \addConstraint{\text{(\ref{PF1CST1}), ~(\ref{PF1CST2}), ~(\ref{PF1CST3}).}\nonumber}
\end{maxi!}
\let\bodyobjLong\oldbodyobjLong
\endgroup

Nonetheless, the reformulated problem $\mathscr{P}_2$ remains non-convex due to the constraints (\ref{PF1CST1}) and (\ref{PF2CST2}), as the spectral efficiency equations are non-convex functions.
To overcome this issue, we propose a WMMSE-based method. From the equation (\ref{common rate_tot}), we observe that the common spectral efficiency can be equivalently derived by decoding the common stream $s_{\sf{c}}$ from the signal $\Vert \mathbf{h}_k \Vert \sqrt{\frac{P_{\sf{c}}}{2}}s_{{\sf{c}}} + \sum_{j=1}^{K}\mathbf{h}_{k}^{\sf{H}}\mathbf{p}_{j}s_{j} + n_{k}$. Similarly, the equation (\ref{private rate_tot}) shows that the private spectral efficiency can be equivalently derived by decoding the private stream $s_{k}$ from the signal $\sum_{j=1}^{K}\mathbf{h}_{k}^{\sf{H}}\mathbf{p}_{j}s_{j} + n_{k}$. Building on these observations, we formulate the common and private mean square errors (MSEs) for the $k$-th user, denoted as $\epsilon_{{\sf{c}},k}$ and $\epsilon_{{\sf{p}}, k}$, with corresponding equalizers $g_{{\sf{c}}, k}$ and $g_{{\sf{p}}, k}$ as follows:
{
\begin{align}
\label{MSE_common}
   \epsilon_{{\sf{c}}, k}  & =  \mathbb{E}\bigg[ \big\vert g_{{\sf{c}}, k} \big(\Vert \mathbf{h}_k \Vert \sqrt{\frac{P_{\sf{c}}}{2}}s_{{\sf{c}}} + \sum_{j=1}^{K}\mathbf{h}_{k}^{\sf{H}}\mathbf{p}_{j}s_{j}+n_{k}\big) - s_{{\sf{c}}} \big\vert^{2}\bigg] \nonumber \\
   & = \vert g_{{\sf{c}}, k} \vert^{2} T_{{\sf{c}}, k} - 2{\sf{Re}}\bigg\{g_{{\sf{c}}, k}\Vert \mathbf{h}_k \Vert \sqrt{\frac{P_{\sf{c}}}{2}}\bigg\} + 1, 
   \\
\label{MSE_private}
   \epsilon_{{\sf{p}}, k} & 
   = \mathbb{E}\bigg[ \big\vert g_{{\sf{p}}, k} \big(\sum_{j=1}^{K}\mathbf{h}_{k}^{\sf{H}}\mathbf{p}_{j}s_{j}+n_{k}\big) - s_{k} \big\vert^{2}\bigg] \nonumber \\
   & = \vert g_{{\sf{p}}, k} \vert^{2} T_{{\sf{p}}, k} - 2{\sf{Re}}\left\{g_{{\sf{p}}, k}\mathbf{h}_k^{\sf{H}}\mathbf{p}_{k}\right\} + 1.   
\end{align}
}Herein, $T_{{\sf{c}}, k}$ and $T_{{\sf{p}}, k}$ are defined as
    $T_{{\sf{c}}, k} \triangleq \Vert \mathbf{h}_k \Vert^{2} {P_{\sf{c}}/2} + \sum_{j=1}^{K} 
    \vert \mathbf{h}_{k}^{\sf{H}} \mathbf{p}_{j}\vert^{2} + \sigma_{{\sf{n}}}^{2}$ and  
    $T_{{\sf{p}}, k} \triangleq \sum_{j=1}^{K} 
    \vert \mathbf{h}_{k}^{\sf{H}} \mathbf{p}_{j}\vert^{2} + \sigma_{{\sf{n}}}^{2}$,
respectively.
With the first-order optimality condition, i.e., $\frac{\partial \epsilon_{{\sf{c}}, k}}{\partial g_{{\sf{c}}, k}} = \frac{\partial \epsilon_{{\sf{p}}, k}}{\partial g_{{\sf{p}}, k}} = 0$, the minimum values of (\ref{MSE_common}) and (\ref{MSE_private}) are obtained when the equalizers are set to $g_{{\sf{c}}, k}^{\sf{MMSE}} = T_{{\sf{c}}, k}^{-1} \Vert \mathbf{h}_k \Vert \sqrt{P_{\sf{c}}/2}$ and $g_{{\sf{p}}, k}^{\sf{MMSE}} = T_{{\sf{p}}, k}^{-1} \mathbf{p}_{k}^{\sf{H}} \mathbf{h}_{k}$, respectively. By substituting $g_{{\sf{c}}, k}^{\sf{MMSE}}$ and $g_{{\sf{p}}, k}^{\sf{MMSE}}$ into (\ref{MSE_common}) and (\ref{MSE_private}), we obtain the common and private minimum MSEs (MMSEs) as $\epsilon_{{\sf{c}}, k}^{\sf{MMSE}} = T_{{\sf{c}}, k}^{-1}(T_{{\sf{c}}, k} - \Vert \mathbf{h}_k \Vert^{2} {P_{\sf{c}}/2})$ and $\epsilon_{{\sf{p}}, k}^{\sf{MMSE}} = T_{{\sf{p}}, k}^{-1}(T_{{\sf{p}}, k} - | \mathbf{h}_{k}^{\sf{H}} \mathbf{p}_{k} |^{2})$, respectively.

Subsequently, by introducing the positive real weights $u_{{\sf{c}}, k}$ and $u_{{\sf{p}}, k}$, we formulate the common and private augmented weighted mean square errors (WMSEs) as follows:
{
\begin{align}
    \label{WMSE}
    \xi_{{\sf{c}}, k} = u_{{\sf{c}}, k} \epsilon_{{\sf{c}}, k} \!-\! \log_2{u_{{\sf{c}}, k}}, \,\, \xi_{{\sf{p}}, k} = u_{{\sf{p}}, k} \epsilon_{{\sf{p}}, k} \!-\! \log_2{u_{{\sf{p}}, k}}.
\end{align}
}According to the first-order optimality condition, the optimum equalizers and weights, which minimize the common WMSE, are obtained by solving  
$\frac{\partial \xi_{{\sf{c}}, k}}{\partial g_{{\sf{c}}, k}} = \frac{\partial \xi_{{\sf{c}}, k}}{\partial u_{{\sf{c}}, k}} = 0$.  
Consequently, the optimum equalizers and weights are given as follows:
{
\begin{align}
\label{common_WMMSE}
g_{{\sf{c}}, k}^{\star} = g_{{\sf{c}}, k}^{\sf{MMSE}}, \,\, u_{{\sf{c}}, k}^{\star} = 1/\epsilon_{{\sf{c}}, k}^{\sf{MMSE}}.
\end{align}
}Similarly, solving $\frac{\partial \xi_{{\sf{p}}, k}}{\partial g_{{\sf{p}}, k}} = \frac{\partial \xi_{{\sf{p}}, k}}{\partial u_{{\sf{p}}, k}} = 0$ yields the optimum equalizers and weights that minimize the private WMSE as
{
\begin{align}
\label{private_WMMSE}
g_{{\sf{p}}, k}^{\star} = g_{{\sf{p}}, k}^{\sf{MMSE}}, \,\, u_{{\sf{p}}, k}^{\star} = 1/\epsilon_{{\sf{p}}, k}^{\sf{MMSE}}.
\end{align}
}Then, by substituting the optimum values (\ref{common_WMMSE}) and (\ref{private_WMMSE}) to the WMSE equations in (\ref{WMSE}), the relationships between WMMSE and spectral efficiency equations are established as follows:
{
\begin{align}
    \label{common_rate_relationship}
       &\xi_{{\sf{c}}, k}^{\star} = \min_{g_{{\sf{c}}, k}, u_{{\sf{c}}, k}} {\xi_{{\sf{c}}, k}} = 1 + \log_2{\epsilon_{{\sf{c}},k}^{\sf{MMSE}}} = 1-R_{{\sf{c}}, k}, \\
    \label{private_rate_relationship}
      &\xi_{{\sf{p}}, k}^{\star} = \min_{g_{{\sf{p}}, k}, u_{{\sf{p}}, k}} {\xi_{{\sf{p}}, k}} = 1 + \log_2{\epsilon_{{\sf{p}},k}^{\sf{MMSE}}} = 1-R_{{\sf{p}}, k}.
\end{align}
}Based on the above relationships, we construct $S$ independent $(\xi_{{\sf{c}},k}^{\star})^{\left(s\right)}$ and $(\xi_{{\sf{p}},k}^{\star})^{\left(s\right)}$ by generating $\mathbf{h}_{k}^{\left(s\right)} = \hat{\mathbf{h}}_{k} + \mathbf{e}_{k}^{\left(s\right)}$, $\forall s \in \mathcal{S}$. By doing so, the equations (\ref{empirical average common rate}) and (\ref{empirical average private rate}) are rewritten as follows:
{
\begin{align}
   \label{empirical common average}
    &\bar{R}_{{\sf{c}},k} = 1 - \bar{\xi}_{{\sf{c}},k}^{\star} = 1 - \frac{1}{S}\sum\limits_{s=1}^{S} (\xi_{{\sf{c}},k}^{\star})^{\left(s\right)}, \\
    \label{empirical private average}
    &\bar{R}_{{\sf{p}},k} = 1 - \bar{\xi}_{{\sf{p}},k}^{\star} = 1 - \frac{1}{S}\sum\limits_{s=1}^{S} (\xi_{{\sf{p}},k}^{\star})^{\left(s\right)}.
\end{align}
}Consequently, $\mathscr{P}_{2}$ is reformulated as $\mathscr{P}_{3}$ since the inequalities $\bar{R}_{{\sf{c}}, k} \geq 1 - \bar{\xi}_{{\sf{c}}, k}$ and $\bar{R}_{{\sf{p}}, k} \geq 1 - \bar{\xi}_{{\sf{p}},k}$, where $\bar{\xi}_{{\sf{c}}, k} = 1/S \cdot \sum_{s=1}^{S} \xi_{{\sf{c}},k}^{\left(s\right)}$ and $\bar{\xi}_{{\sf{p}}, k} = 1/S \cdot \sum_{s=1}^{S} \xi_{{\sf{p}},k}^{\left(s\right)}$, always hold.
\begingroup
\let\oldbodyobjLong\bodyobjLong
\renewcommand{\bodyobjLong}[4]{%
  \ifthenelse{\isempty{#4}}{%
    &\text{\normalsize$\mathscr{P}_3:$}~\operatorname*{maximize}_{\displaystyle #1}\ #2\nonumber\span\span\span\span%
  }{%
    #4~&\text{\normalsize$\mathscr{P}_3:$}~\operatorname*{maximize}_{\displaystyle #1}\ #2\nonumber\span\span\span\span%
  }%
}
\begin{maxi!}|l|[2]
  {\substack{P_{\sf{c}},\, \mathbf{P},\, \mathbf{c},\, \boldsymbol{\alpha}_{\sf{p}},\, q, \\ \mathbf{G}_{\sf{c}},\, \mathbf{G}_{\sf{p}},\, \mathbf{U}_{\sf{c}},\, \mathbf{U}_{\sf{p}}}}
  {q}
  {}{} 
  \addConstraint{1 - \bar{\xi}_{{\sf{p}}, k} \geq \alpha_{{\sf{p}}, k}, ~\forall k \in \mathcal{K},}\label{WMMSECST1}
  \addConstraint{1 - \bar{\xi}_{{\sf{c}},k} \geq \sum_{j=1}^{K}C_{j}, ~\forall k \in \mathcal{K},}\label{WMMSECST2}
  \addConstraint{\text{(\ref{PF1CST2})}, ~\textrm{(\ref{PF1CST3})}, ~\textrm{(\ref{PF2CST1})},  ~\textrm{(\ref{PF2CST3}).}\nonumber}
\end{maxi!}
\let\bodyobjLong\oldbodyobjLong
\endgroup
Herein, $\mathbf{G}_{\sf{c}}$ and $\mathbf{G}_{\sf{p}}$ are defined as $\mathbf{G}_{\sf{c}}\triangleq\{\mathbf{g}_{{\sf{c}}, k} \, \vert \, \forall k \in \mathcal{K}\}$ and $\mathbf{G}_{\sf{p}}\triangleq\{\mathbf{g}_{{\sf{p}}, k} \, \vert \, \forall k \in \mathcal{K}\}$, respectively, where $\mathbf{g}_{{\sf{c}}, k} \triangleq \{g_{{\sf{c}},k}^{(s)} \, \vert \, \forall s \in \mathcal{S}\}$ and $\mathbf{g}_{{\sf{p}}, k} \triangleq \{g_{{\sf{p}},k}^{(s)} \, \vert \, \forall s \in \mathcal{S}\}$. Similarly, $\mathbf{U}_{\sf{c}}$ and $\mathbf{U}_{\sf{p}}$ are defined as $\mathbf{U}_{\sf{c}}\triangleq\{\mathbf{u}_{{\sf{c}}, k} \, \vert \,  \forall k \in \mathcal{K}\}$ and $\mathbf{U}_{\sf{p}}\triangleq\{\mathbf{u}_{{\sf{p}}, k} \, \vert \, \forall k \in \mathcal{K}\}$, {respectively}, where $\mathbf{u}_{{\sf{c}}, k} \triangleq \{u_{{\sf{c}},k}^{(s)} \, \vert \, \forall s \in \mathcal{S}\}$ and $\mathbf{u}_{{\sf{p}}, k} \triangleq \{u_{{\sf{p}},k}^{(s)} \, \vert \, \forall s \in \mathcal{S}\}$.  
Although $\mathscr{P}_{3}$ is non-convex with respect to the joint variable set, we tackle it through an alternating optimization approach that iteratively optimizes a subset of variables while fixing the rest, as follows:


{\textbf{Step I}:} At the $n$-th iteration, $\mathbf{G}_{\sf{c}}^{[n]}$, $\mathbf{G}_{\sf{p}}^{[n]}$, $\mathbf{U}_{\sf{c}}^{[n]}$, and $\mathbf{U}_{\sf{p}}^{[n]}$ are first updated based on equations (\ref{common_WMMSE}) and (\ref{private_WMMSE}), using $P_{\sf{c}}^{[n-1]}$ and $\mathbf{P}^{[n-1]}$ that are obtained from the previous iteration.
Then, $\forall k \in \mathcal{K}$ and $\forall s \in \mathcal{S}$, the following variables are calculated based on the updated $\mathbf{G}_{\sf{c}}^{[n]}$, $\mathbf{G}_{\sf{p}}^{[n]}$, $\mathbf{U}_{\sf{c}}^{[n]}$, and $\mathbf{U}_{\sf{p}}^{[n]}$.
{
\begin{align}
 \label{step1_1}
    \tau^{(s)}_{{\sf{c}},k} = u^{(s)}_{{\sf{c}},k} \vert g^{(s)}_{{\sf{c}},k} \vert^2, \,\, \tau^{(s)}_{{\sf{p}},k} = u^{(s)}_{{\sf{p}},k} \vert g^{(s)}_{{\sf{p}},k} \vert^2, \,\, {\psi}^{(s)}_{{\sf{c}},k} = \tau^{(s)}_{{\sf{c}},k}  \Vert \mathbf{h}^{(s)}_{k} \Vert^{2},
\end{align}
\begin{align}
 \label{step1_3}
    \boldsymbol{\Psi}^{(s)}_{{\sf{c}},k} = \tau^{(s)}_{{\sf{c}},k} \mathbf{h}^{(s)}_{k} (\mathbf{h}^{(s)}_{k})^{\sf{H}}, \,\, \boldsymbol{\Psi}^{(s)}_{{\sf{p}},k} = \tau^{(s)}_{{\sf{p}},k} \mathbf{h}^{(s)}_{k} (\mathbf{h}^{(s)}_{k})^{\sf{H}}, 
\end{align}
\begin{align}
 \label{step1_4}
    w^{(s)}_{{\sf{c}},k} = u^{(s)}_{{\sf{c}},k} g^{(s)}_{{\sf{c}},k} \Vert \mathbf{h}^{(s)}_{k} \Vert, \,\, \mathbf{w}^{(s)}_{{\sf{p}},k} = u^{(s)}_{{\sf{p}},k} (g^{(s)}_{{\sf{p}},k})^{*} \mathbf{h}^{(s)}_{k},
\end{align}
\begin{align}
 \label{step1_5}
    v^{(s)}_{{\sf{c}},k} = \log_2 u^{(s)}_{{\sf{c}},k}, \,\, v^{(s)}_{{\sf{p}},k} = \log_2 u^{(s)}_{{\sf{p}},k}.
\end{align}
}Following this, the sample-averaged values, denoted as $\bar{\tau}_{{\sf{c}},k}$, $\bar{\tau}_{{\sf{p}},k}$, $\bar{\psi}_{{\sf{c}},k}$, $\bar{\boldsymbol{\Psi}}_{{\sf{c}},k}$, $\bar{\boldsymbol{\Psi}}_{{\sf{p}},k}$, $\bar{w}_{{\sf{c}},k}$, $\bar{\mathbf{w}}_{{\sf{p}},k}$, $\bar{v}_{{\sf{c}},k}$, $\bar{v}_{{\sf{p}},k}$, $\bar{u}_{{\sf{c}},k}$, and $\bar{u}_{{\sf{p}},k}$ are obtained by taking averages over $S$ realizations.

{\textbf{Step II}:} In this step, we update $P_{\sf{c}}^{[n]}$, $\mathbf{P}^{[n]}$, $\mathbf{c}^{[n]}$, $\boldsymbol{\alpha}_{\sf{p}}^{[n]}$, and $q^{[n]}$ based on the updated values of $\mathbf{G}_{\sf{c}}^{[n]}$, $\mathbf{G}_{\sf{p}}^{[n]}$, $\mathbf{U}_{\sf{c}}^{[n]}$, and $\mathbf{U}_{\sf{p}}^{[n]}$ from \textbf{Step I}. To be specific, $\bar{\xi}_{{\sf{c}}, k}$ and $\bar{\xi}_{{\sf{p}}, k}$ in $\mathscr{P}_3$ are substituted based on the sample-averaged values $\bar{\tau}_{{\sf{c}},k}$, $\bar{\tau}_{{\sf{p}},k}$, $\bar{\psi}_{{\sf{c}},k}$, $\bar{\boldsymbol{\Psi}}_{{\sf{c}},k}$, $\bar{\boldsymbol{\Psi}}_{{\sf{p}},k}$, $\bar{w}_{{\sf{c}},k}$, $\bar{\mathbf{w}}_{{\sf{p}},k}$, $\bar{v}_{{\sf{c}},k}$, $\bar{v}_{{\sf{p}},k}$, $\bar{u}_{{\sf{c}},k}$, and $\bar{u}_{{\sf{p}},k}$, which are obtained using $\mathbf{G}_{\sf{c}}^{[n]}$, $\mathbf{G}_{\sf{p}}^{[n]}$, $\mathbf{U}_{\sf{c}}^{[n]}$, and $\mathbf{U}_{\sf{p}}^{[n]}$.
Thereafter, $P_{\sf{c}}^{[n]}$, $\mathbf{P}^{[n]}$, $\mathbf{c}^{[n]}$, $\boldsymbol{\alpha}_{\sf{p}}^{[n]}$, and $q^{[n]}$ are derived by solving the following convex optimization problem based on the interior-point method.
\begingroup
\let\oldbodyobjLong\bodyobjLong
\renewcommand{\bodyobjLong}[4]{%
  \ifthenelse{\isempty{#4}}{%
    &\text{\normalsize$\mathscr{P}_3^{[n]}:$}~\operatorname*{maximize}_{\displaystyle #1}\ #2\nonumber\span\span\span\span%
  }{%
    #4~&\text{\normalsize$\mathscr{P}_3^{[n]}:$}~\operatorname*{maximize}_{\displaystyle #1}\ #2\nonumber\span\span\span\span%
  }%
}
\begin{maxi!}|l|[2]
  {_{P_{\sf c},\, \mathbf{P},\, \mathbf{c},\, \boldsymbol{\alpha}_{\sf p},\, q}}
  {q}
  {}{} 
  \addConstraint{1 - \sum_{j=1}^{K}\mathbf{p}_{j}^{\sf{H}}\bar{\boldsymbol{\Psi}}_{{\sf{p}}, k}\mathbf{p}_{j}  +2{\sf{Re}}\{\bar{\mathbf{w}}^{\sf{H}}_{{\sf{p}},k}\mathbf{p}_{k}\} \nonumber}
  \addConstraint{- \bar{\tau}_{{\sf{p}},k}\sigma_{\sf{n}}^{2} - \bar{u}_{{\sf{p}},k} + \bar{v}_{{\sf{p}},k} \geq \alpha_{{\sf{p}}, k},  ~\forall k \in \mathcal{K},}\label{WMMSECST1_n}
  \addConstraint{1 - \bar{\psi}_{{\sf c},k}\frac{P_{\sf c}}{2}
    - \sum_{j=1}^{K}\mathbf{p}_{j}^{\sf H}\bar{\boldsymbol{\Psi}}_{{\sf c},k}\mathbf{p}_{j}
    + 2{\sf Re}\!\left\{\bar{w}_{{\sf c},k}\sqrt{\frac{P_{\sf c}}{2}}\right\} \nonumber}
  \addConstraint{- \bar{\tau}_{{\sf c},k}\sigma_{\sf n}^{2} - \bar{u}_{{\sf c},k} + \bar{v}_{{\sf c},k} \ge \sum_{j=1}^{K}C_j,~\forall k\in\mathcal{K},}\label{WMMSECST2_n}
  \addConstraint{\text{(\ref{PF1CST2})}, ~\textrm{(\ref{PF1CST3})}, ~ \textrm{(\ref{PF2CST1})}, ~ \textrm{(\ref{PF2CST3}).}\nonumber}
\end{maxi!}
\let\bodyobjLong\oldbodyobjLong
\endgroup

These steps are repeated until the absolute difference between consecutive objective function values, $\vert q^{[n]} - q^{[n-1]}\vert$, falls below a predefined threshold $\epsilon$. {The detailed procedure is summarized in \textbf{Algorithm \ref{Algorithm 1}}.}
Notably, the solution of $\mathscr{P}_{3}^{[n-1]}$ remains within the feasible set of $\mathscr{P}_{3}^{[n]}$, ensuring that $q^{[n]}$ forms a non-decreasing sequence as iterations progress. Furthermore, the objective function is upper-bounded by the given total transmit power constraint, which, together with the algorithm's structure, ensures guaranteed convergence. 

{The primary computational complexity of WMMSE-based ST-RSMA precoder design arises from solving the convex optimization problem in \textbf{Step II}.
Since the problem $\mathscr{P}_{3}^{[n]}$ can be efficiently solved by the interior-point method, the computational complexity of the proposed ST-RSMA framework is characterized in big-O notation as $\mathcal{O}\big([N_{\sf{t}}K]^{3.5} \log(\epsilon^{-1})\big)$.}



\section{Extension to Increased Number of Antennas}
In this section, we extend the proposed ST-RSMA-based precoder design to the scenarios where the LEO satellite is equipped with more than two transmit antenna feeds, i.e., $N_{\sf{t}} > 2$. To this end, we define the set of transmit antenna feed pairs as 
{ $\mathcal{A} = \{(m,n) \vert 1 \leq m < n \leq N_{\sf{t}}\}$}
in which $\mathcal{A}$ consists of $\binom{N_{\sf{t}}}{2}$ distinct pairs.
Subsequently, we evaluate the quality of the channel $\mathbf{h}_{k,(m,n)} \in \mathbb{C}^{2\times1}$, which corresponds to the $(m,n)$-th antenna feed pair from $\mathbf{h}_{k} \in \mathbb{C}^{N_{\sf t}\times1}$.
To elaborate further, since the common spectral efficiency is constrained by the user with the lowest common spectral efficiency, we compute  
{ $ \mathbb{E}[\Vert \mathbf{h}_{k,(m,n)} \Vert^{2}] = \mathbb{E}[\Vert \hat{\mathbf{h}}_{k,(m,n)} + \mathbf{e}_{k,(m,n)} \Vert^{2}] = \Vert \hat{\mathbf{h}}_{k,(m,n)} \Vert^{2} + \mathbb{E}[\Vert\mathbf{e}_{k,(m,n)}\Vert^{2}]$} for all $k \in \mathcal{K}$, and then determine
$ \min_{k \in \mathcal{K}} \mathbb{E}[\Vert \mathbf{h}_{k,(m,n)} \Vert^{2}]$.
Lastly, we select the optimal transmit antenna feed pair that satisfies 
$(m^{\star}, n^{\star}) = \argmax_{(m,n) \in \mathcal{A}} \min_{k \in \mathcal{K}} \mathbb{E}[\Vert \mathbf{h}_{k, (m,n)} \Vert^{2}]$.

We then construct a matrix $\mathbf{\Pi}\in \mathbb{R}^{N_{\sf{t}}\times2}$ whose elements are 
{
\begin{align}
\Pi_{i,j} =
\begin{cases}
1, & \text{if} \, (i = m^{\star} \text{ and } j = 1) \,\, \text{or} \,\, (i = n^{\star} \text{ and }  j = 2), \\
0, & \text{otherwise}. 
\end{cases}  
\end{align}
}The LEO satellite linearly combines $\mathbf{s}_{ {\sf{c}}}^{(1)}$ and $\mathbf{s}_{ {\sf{c}}}^{(2)}$ with the constructed matrix $\mathbf{\Pi}$, and then sequentially transmits the signal over consecutive symbol periods as { $\mathbf{x}(t) = \sqrt{\frac{P_{\sf{c}}}{2}} \mathbf{\Pi}\mathbf{s}_{ {\sf{c}}}^{(1)} + \sum_{j=1}^{K} \mathbf{p}_{j} s_{j}^{(1)} \in \mathbb{C}^{N_{\sf{t}}\times1}$} and  { $\mathbf{x}(t+T) = \sqrt{\frac{P_{\sf{c}}}{2}} \mathbf{\Pi}\mathbf{s}_{ {\sf{c}}}^{(2)} + \sum_{j=1}^{K} \mathbf{p}_{j} s_{j}^{(2)}\in \mathbb{C}^{N_{\sf{t}}\times1}$}.
Subsequently, the common and private spectral efficiencies are derived as in the following equations by applying the same combining and decoding processes at the receiver, demonstrated in Section \uppercase\expandafter{\romannumeral2}.
\begin{align}
     &{R}_{{\sf{c}},k} \!=\! \sum_{i=1}^{2}{R}_{{\sf{c}},k}^{(i)} = \log_2\left(1+\frac{\Vert\mathbf{h}_{k,(m^{\star}, n^{\star})}\Vert^2 \frac{P_{\sf{c}}}{2}}{\sum_{j=1}^{K}\vert\mathbf{h}_{k}^{\sf{H}}\mathbf{p}_{j}\vert^{2} + \sigma_{{\sf{n}}}^{2}}\right), \\
     &{R}_{{\sf{p}},k} \!=\! \sum_{i=1}^{2}{R}_{{\sf{p}},k}^{(i)} = \log_2\left(\!1\!+\!\frac{\vert\mathbf{h}_{k}^{\sf{H}}\mathbf{p}_{k}\vert^{2}}{\sum_{j=1, j \neq k }^{K}\vert\mathbf{h}_{k}^{\sf{H}}\mathbf{p}_{j}\vert^{2} \!+\! \sigma_{{\sf{n}}}^{2}}\!\right)\! 
\end{align}
Then, by following the same optimization procedure, outlined in Section \uppercase\expandafter{\romannumeral3}, an optimal solution can be efficiently obtained.

\begin{algorithm}[!t]{
\caption{WMMSE-Based ST-RSMA Precoder Design}\label{Algorithm 1}
{ \begin{algorithmic}[1]
\State \textbf{Input}: $P_{\sf{t}}$, $N_{\sf{t}}$, $K$, $\sigma_{\sf{n}}^{2}$, $S$, $\epsilon$, $\hat{\mathbf{h}}_{k}$, and $\mathbf{\Phi}_{k}$, $\forall k \in \mathcal{K}$.
\State \textbf{Initialize}: $P_{\sf{c}}^{[0]}$, $\mathbf{P}^{[0]}$, $q^{[0]}$, and $n \leftarrow 0$.
\Repeat
      \State $n \leftarrow n+1$.
      \State Update $\mathbf{G}_{\sf{c}}^{[n]}$, $\mathbf{G}_{\sf{p}}^{[n]}$, $\mathbf{U}_{\sf{c}}^{[n]}$, and $\mathbf{U}_{\sf{p}}^{[n]}$ based on (\ref{common_WMMSE}) and (\ref{private_WMMSE}).
      \State Compute the variables (\ref{step1_1})$-$(\ref{step1_5}), $\forall k \in \mathcal{K}$ and $\forall s \in \mathcal{S}$, using updated $\mathbf{G}_{\sf{c}}^{[n]}$,  $\mathbf{G}_{\sf{p}}^{[n]}$, $\mathbf{U}_{\sf{c}}^{[n]}$, and $\mathbf{U}_{\sf{p}}^{[n]}$.
      \State Solve $\mathscr{P}_{3}^{[n]}$ and obtain $P_{\sf{c}}^{[n]}$, $\mathbf{P}^{[n]}$,  $\mathbf{c}^{[n]}$, $\boldsymbol{\alpha}_{\sf{p}}^{[n]}$, and $q^{[n]}$.
\Until{$\vert q^{[n]} - q^{[n-1]}\vert \leq \epsilon$.}
\State \textbf{Output}: $P_{\sf{c}}^{[n]}$, $\mathbf{P}^{[n]}$, and $\mathbf{c}^{[n]}$.
\end{algorithmic}}}
\end{algorithm}


\section{Performance Evaluation} \label{Sec4}
In this section, we evaluate the performance of the proposed ST-RSMA framework. {The simulation parameters are configured based on the 3GPP NTN standards \cite{3gpp_ntn}.} The LEO satellite operates at an altitude of $\num{600}$ km, with each spot beam covering a radius of $\num{25}$ km, corresponding to a $3$ dB beamwidth of $4.4127^{\circ}$. The carrier frequency, bandwidth, maximum transmit antenna gain, user antenna gain, receiver noise temperature, noise variance, number of SAA samples, and tolerance value are set to $f_{\sf{c}}=\num{20}$ GHz, $B=\num{400}$ MHz, $G_{\sf{Tx}}^{\sf{max}} = \num{30.5}$ dBi, $G_{\sf{Rx}} = \num{39.7}$ dBi, $T_{\sf{sys}} = \num{150}$ K, $\sigma_{\sf{n}}^{2} = \num{1}$, $S = \num{1000}$, and $\epsilon = \num{10}^{-4}$, respectively. The CSI error vector $\mathbf{e}_k$ is set to follow $\mathbf{e}_k\sim\mathcal{CN}(\mathbf{0}, \sigma_{{\sf{e}},k}^2 \mathbf{I})$, where the same variance of $\sigma_{\sf{e}}^2$ is considered among all users. {The total transmit power budget at the LEO satellite is set to $P_{\sf{t}} = 30$ dBm, unless mentioned otherwise. 
For benchmarks, we consider several conventional RSMA schemes, including WMMSE-based RSMA \cite{yin2020rate_J} (``WMMSE-RSMA'') and successive convex approximation (SCA)-based RSMA \cite{cui2023energy} (``SCA-RSMA''), as well as baseline MA techniques, such as SDMA, multicasting, and fractional resource reuse where orthogonal resources are allocated to each spot beam.}\footnote{{To account for the effects of imperfect CSIT in the design of the SCA-RSMA precoder, we employ the generalized mutual information technique.}} {In all RSMA-based schemes, including ST-RSMA, the LEO satellite dynamically reallocates the common portions based on users' channel quality indicator (CQI) report to ensure fair rates, requiring only minimal signaling.}
Numerical results are averaged over $\num{1000}$ independent channel realizations with randomly placed users.

\begin{figure*}[!t]
\centering
\includegraphics[width=1\linewidth]{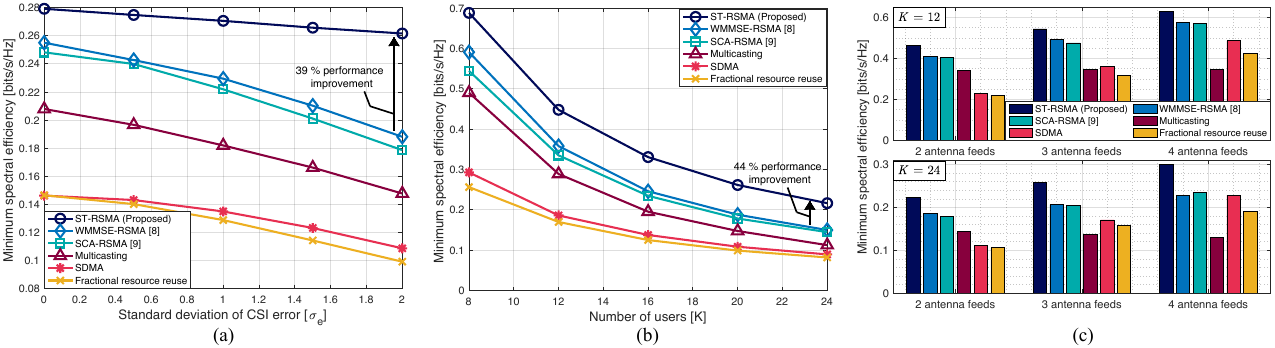}
\caption{{Minimum spectral efficiency comparisons under various system settings: (a) As a function of 
$\sigma_{\sf{e}}$ with $N_{\sf{t}}$ = 2 and $K$ = 20; (b) As a function of 
$K$ with $N_{\sf{t}}$ = 2 and $\sigma_{\sf{e}}$ = 2; (c) As a function of 
$N_{\sf{t}}$ with $\sigma_{\sf{e}}$ = 1, for $K$ = 12 (upper subplot) and $K$ = 24 (lower subplot).}}
\label{Fig_sim_1} 
\end{figure*}

Fig.~\ref{Fig_sim_1}(a) compares the minimum spectral efficiency as a function of the standard deviation of CSI error $ \sigma_{\sf{e}} $. The number of transmit antenna feeds and users are set to $ N_{\sf{t}}=2 $ and $ K=20 $, respectively. ST-RSMA consistently outperforms existing MA techniques across all $ \sigma_{\sf{e}} $ regimes. The performance gap between ST-RSMA and other techniques widens as $ \sigma_{\sf{e}} $ increases, highlighting the robustness of ST-RSMA scheme against imperfect CSIT. {Specifically, at $ \sigma_{\sf{e}}=2 $, ST-RSMA achieves a \num{39}\% performance gain over WMMSE-RSMA.}

Fig.~\ref{Fig_sim_1}(b) evaluates the minimum spectral efficiency as a function of the number of users $K$ when $N_{\sf{t}}=2$ and $\sigma_{\sf{e}}=2$. The results indicate that ST-RSMA exhibits an increase in robustness over other MA techniques as the number of users grows, highlighting its effectiveness in network scalability. {The performance gap between ST-RSMA and WMMSE-RSMA is \num{16}\% when $K=8$, whereas it expands to \num{44}\% when $K=24$.}

In Fig.~\ref{Fig_sim_1}(c), we evaluate the minimum spectral efficiency as a function of the number of transmit antenna feeds $N_{\sf{t}}$ for 12 and 24 users when $\sigma_{\sf{e}}=1$. Regardless of the number of transmit antenna feeds, ST-RSMA consistently outperforms other MA techniques in both user scenarios. {In particular, the performance gap between ST-RSMA and other MA techniques continues to widen as the number of users increases, irrespective of the number of transmit antenna feeds, highlighting its scalability. For instance, ST-RSMA outperforms WMMSE-RSMA by \num{10}\% and \num{9}\% for 12 users when $N_{\sf{t}}=3$ and $N_{\sf{t}}=4$, respectively; these gains increase to \num{24}\% and \num{31}\% for 24 users.}


\begin{figure*}[!t]
\centering
\includegraphics[width=1\linewidth]{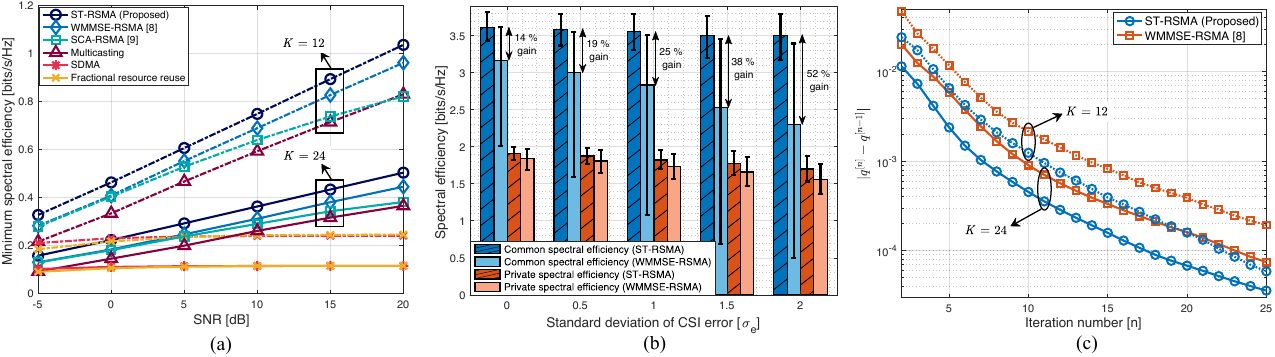}
\caption{{(a) Minimum spectral efficiency comparison as a function of SNR with $N_{\sf{t}}=2$ and $\sigma_{\sf{e}}=1$, for $K$ = 12 and $K$ = 24; (b) Achievable spectral efficiency comparison as a function of $\sigma_{\sf{e}}$ with $N_{\sf{t}}$ = 2 and $K$ = 24;  (c) Convergence comparison under $N_{\sf{t}} = 2$ and $\sigma_{\sf{e}}$ = 1, for $K$ = 12 and $K$ = 24.}}
\label{Fig_sim_2} 
\end{figure*}

{Fig.~\ref{Fig_sim_2}(a) compares the minimum spectral efficiency as a function of the signal-to-noise ratio (SNR) for $N_{\sf{t}} = 2$ and $\sigma_{\sf{e}} = 1$, with $K = 12$ and $K = 24$. It is shown that ST-RSMA consistently outperforms existing MA techniques across entire SNR regimes for both cases, demonstrating its effectiveness regardless of the satellite's onboard transmit power availability.}

{To quantitatively highlight the space-time diversity gain of ST-RSMA for the common stream transmission, Fig.~\ref{Fig_sim_2}(b) presents the minimum common spectral efficiency (i.e., { $\min\limits_{k \in \mathcal{K}} R_{{\sf{c}},k}$}) and the total private spectral efficiency (i.e., { $\sum_{j=1}^{K} R_{{\sf{p}},j}$}) as a function of $ \sigma_{\sf{e}} $, with $ N_{\sf{t}} = 2 $ and $ K = 24 $. The black lines on each bar represent the range of corresponding values across channel realizations. While the total private spectral efficiency remains comparable to that of conventional RSMA, ST-RSMA achieves a significantly higher minimum common spectral efficiency with reduced variance. These results demonstrate that the space-time diversity gain of the proposed ST-RSMA enhances system robustness and flexibility by providing a reliably higher minimum common spectral efficiency compared to conventional RSMA across diverse network conditions.}

{Fig.~\ref{Fig_sim_2}(c) illustrates the convergence behavior of ST-RSMA compared to WMMSE-RSMA. By eliminating the need for a common beamforming vector design, ST-RSMA achieves faster convergence and lower computational overhead. These results validate the effectiveness of ST-RSMA in achieving improved performance with reduced computational cost.}



{\section{Conclusion and Future Directions}} 
\label{Sec4}
This paper proposed the ST-RSMA framework, which incorporates space-time coding into the common stream transmission. {The proposed approach provides full diversity gain for the common stream across all users, regardless of channel conditions and network load, and has been shown to significantly enhance the minimum spectral efficiency compared to conventional RSMA and other MA techniques.}

{As a future direction, it is promising to consider hardware non-idealities, such as RF chain impairment, in the ST-RSMA precoder design, particularly given the limited feasibility of hardware repair in satellite systems. 
Developing ST-RSMA schemes that are robust to such impairments will be crucial for ensuring resilient and long-term LEO SATCOM services.}
\vspace{5mm}

\bibliographystyle{IEEEtran}
\bibliography{jhseong_reff}

\begin{thebibliography}{10}
\providecommand{\url}[1]{#1}
\csname url@samestyle\endcsname
\providecommand{\newblock}{\relax}
\providecommand{\bibinfo}[2]{#2}
\providecommand{\BIBentrySTDinterwordspacing}{\spaceskip=0pt\relax}
\providecommand{\BIBentryALTinterwordstretchfactor}{4}
\providecommand{\BIBentryALTinterwordspacing}{\spaceskip=\fontdimen2\font plus
\BIBentryALTinterwordstretchfactor\fontdimen3\font minus \fontdimen4\font\relax}
\providecommand{\BIBforeignlanguage}[2]{{%
\expandafter\ifx\csname l@#1\endcsname\relax
\typeout{** WARNING: IEEEtran.bst: No hyphenation pattern has been}%
\typeout{** loaded for the language `#1'. Using the pattern for}%
\typeout{** the default language instead.}%
\else
\language=\csname l@#1\endcsname
\fi
#2}}
\providecommand{\BIBdecl}{\relax}
\BIBdecl

\bibitem{perez2019signal}
A.~I. Perez-Neira \emph{et~al.}, ``Signal processing for high-throughput satellites: {C}hallenges in new interference-limited scenarios,'' \emph{IEEE Signal Process. Mag.}, vol.~36, no.~4, pp. 112--131, 2019.

\bibitem{jamshed2025tutorial}
M.~A. Jamshed \emph{et~al.}, ``A tutorial on non-terrestrial networks: Towards global and ubiquitous {6G} connectivity,'' \emph{Foundations and Trends{\textregistered} in Networking}, vol.~14, no.~3, pp. 160--253, 2025.

\bibitem{3gpp_ntn}
3GPP, ``Solutions for {NR} to support non-terrestrial networks ({NTN}) ({R}elease 16),'' \emph{3GPP TR 38.821}, v16.2.0, 2023.

\bibitem{you2020massive}
L.~You \emph{et~al.}, ``Massive {MIMO} transmission for {LEO} satellite communications,'' \emph{IEEE J. Sel. Areas Commun.}, vol.~38, no.~8, pp. 1851--1865, 2020.

\bibitem{10844052}
Q.~Li \emph{et~al.}, ``Holographic metasurface-based beamforming for multi-altitude {LEO} satellite networks,'' \emph{IEEE Trans. Wireless Commun.}, vol.~24, no.~4, pp. 3103--3116, 2025.

\bibitem{10559954}
M.~Toka \emph{et~al.}, ``{RIS}-empowered {LEO} satellite networks for {6G}: Promising usage scenarios and future directions,'' \emph{IEEE Commun. Mag.}, vol.~62, no.~11, pp. 128--135, 2024.

\bibitem{park2023rate}
J.~Park \emph{et~al.}, ``Rate-splitting multiple access for {6G} networks: Ten promising scenarios and applications,'' \emph{IEEE Network}, vol.~38, no.~3, pp. 128--136, 2024.

\bibitem{yin2020rate_J}
L.~Yin and B.~Clerckx, ``Rate-splitting multiple access for multigroup multicast and multibeam satellite systems,'' \emph{IEEE Trans. Commun.}, vol.~69, no.~2, pp. 976--990, 2020.

\bibitem{cui2023energy}
H.~Cui \emph{et~al.}, ``Energy-efficient {RSMA} for multigroup multicast and multibeam satellite communications,'' \emph{IEEE Wireless Commun. Lett.}, vol.~12, no.~5, pp. 838--842, 2023.

\bibitem{10266774}
J.~Lee \emph{et~al.}, ``Coordinated rate-splitting multiple access for integrated satellite-terrestrial networks with super-common message,'' \emph{IEEE Trans. Veh. Technol.}, vol.~73, no.~2, pp. 2989--2994, 2024.

\bibitem{10636955}
J.~Ryu \emph{et~al.}, ``Rate-splitting multiple access for {GEO}-{LEO} coexisting satellite systems: A traffic-aware throughput maximization precoder design,'' \emph{IEEE Trans. Veh. Technol.}, vol.~73, no.~12, pp. 19\,838--19\,843, 2024.

\bibitem{alamouti1998simple}
S.~M. Alamouti, ``A simple transmit diversity technique for wireless communications,'' \emph{IEEE J. Sel. Areas Commun.}, vol.~16, no.~8, pp. 1451--1458, 1998.

\bibitem{dahlman20205g}
E.~Dahlman, S.~Parkvall, and J.~Skold, \emph{5G NR: The next generation wireless access technology}.\hskip 1em plus 0.5em minus 0.4em\relax Academic Press, 2020.

\end{thebibliography}
\end{document}